# Phonons as a Probe of the Low Temperature Metal Insulator Transition in $Na_{0.75}Co_{0.95}Ni_{0.05}O_2$

**M. Premila*, A. Bharathi, P. Yasodha, N. Gayathri, Y. Hariharan and C. S. Sundar**
Materials Science Division, Indira Gandhi Centre for Atomic Research, Kalpakkam - 603 102, T.N
*Corresponding author: E-mail: premila@igcar.ernet.in

***Abstract***

*Nickel substitution at the cobalt site in $Na_xCoO_2$ induces a Metal Insulator Transition (MIT) and the temperature at which this occurs ($T_{MIT}$) increases with Ni content. Low temperature far infrared measurements on polycrystalline samples of $Na_{0.75}CoO_2$ and $Na_{0.75}Co_{0.95}Ni_{0.05}O_2$ ($T_{MIT}$ = ~ 150K ) were carried out to look for signatures of this transition. While both the samples show an evident splitting of the high frequency Co-O mode on lowering the temperature, dramatic changes are observed in the low frequency sodium mode in $Na_{0.75}Co_{0.95}Ni_{0.05}O_2$ clearly pointing out to an ordering of the sodium ions in the low temperature insulating state.*

**INTRODUCTION**

Layered oxides of the form $Na_xCoO_2$ have attracted the attention of researchers after the recent discovery of superconductivity at $T_c$ ~5K in the hydrated cobaltate $Na_{.3}CoO_2.1.3H_2O$ [1], while the related anhydrous metallic compound $Na_{.75}CoO_2$ exhibits enhanced thermopower [2]. Crystal structures of these compounds involve $CoO_2$ layers formed from edge sharing $CoO_6$ octahedra, alternating with Na ions in two partially occupied sites - Na(1) and Na(2) (Fig.1)[3]. The physical properties in these materials strongly depend on the sodium concentration and for x=0.5 the system undergoes a transition to the insulating state at ~53K. The observed insulating transition is said to be associated to a charge ordering in the $CoO_2$ planes which further induces an ordering of the sodium ions [4]. Substituting the Co sub lattice in $Na_{.75}CoO_2$ with nickel induces drastic changes in its electronic properties and the system undergoes a transition to the insulating state (Fig.1) [5]. Since infrared spectra are sensitive to these local changes in the system associated with the charge ordered metal – insulator transition, it is of immense interest to follow the phonon modes across the metal –insulator transition. Here we report the low temperature (300K-77K) far infrared absorption measurements on polycrystalline samples of $Na_{0.75}CoO_2$ and $Na_{0.75}Co_{0.95}Ni_{0.05}O_2$ ($T_{MIT}$ ~150K).

**EXPERIMENTAL DETAILS**

Polycrystalline samples of $Na_{0.75}CoO_2$ and $Na_{0.75}Co_{0.95}Ni_{0.05}O_2$ were prepared by the conventional solid state route [5]. FTIR absorption measurements were carried out on well characterized powder samples using a Bomem DA8 spectrometer operating with a resolution of 4 $cm^{-1}$. Measurements in the far infrared range were done using a combination of the extended 6μm mylar beamsplitter and the DTGS detector in the range 200 – 750 $cm^{-1}$ on samples pelletised with CsI. Low temperature measurements were carried out on samples mounted inside a JANIS continuous flow cryostat in the temperature range 300 – 77K.

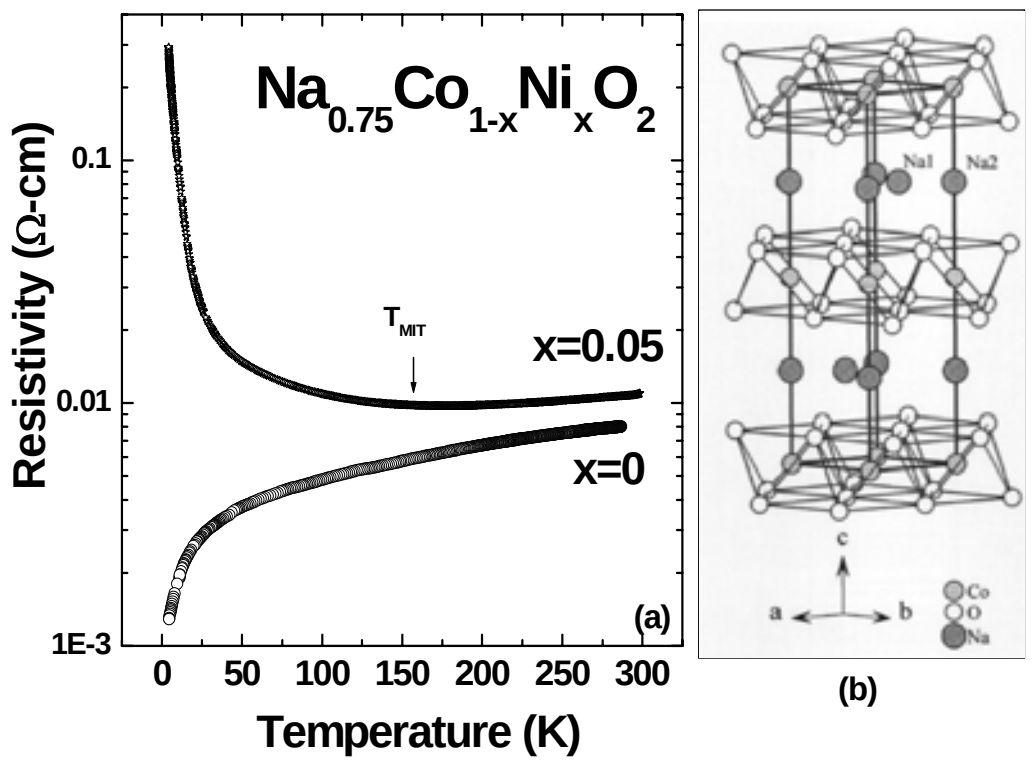

**Fig.1.** (a) Variation of resistivity with temperature. (b) The crystal structure of $Na_xCoO_2$ showing the two sodium sites [3].

**RESULTS AND DISCUSSION**

Room temperature infrared absorption spectra of the undoped $Na_{0.75}CoO_2$ revealed the four expected infrared active modes - the two low frequency modes at 241 and 274 $cm^{-1}$ corresponding to the in plane and out of plane vibration of sodium and a broad high frequency mode at 574 $cm^{-1}$ comprising of both the out of plane and in plane vibrations of cobalt ( Fig.2) [6]. It is to be noted that these low frequency sodium modes have been discerned for the first time and this has been possible because the present measurements have been carried out in the absorption geometry. We try to assign these Na modes based on first principle calculations [6]. The in plane mode is found to occur at lower frequency and is insensitive to Na position, thus the unsplit 241cm$^{-1}$ mode can be assigned to this vibration. The out of plane mode on the other hand is very sensitive to the sodium position [6], with the Na(2) position giving rise to the higher and Na(1) position giving rise to the lower frequency modes. We assign the modes observed at 277 cm$^{-1}$ and 297 cm$^{-1}$ to these vibrations. Neutron scattering data on the sample with $Na_{0.7}CoO_2$ indicates a Na(1):Na(2) occupation ratio of 0.5:0.2. The observed intensity ratios of the Na(2) and Na(1) out of plane vibrations (see Fig.2) are also consistent with these site occupancy ratios. Although no new modes are seen to appear with nickel substitution, significant changes in the phonon modes are seen on lowering the temperature (Fig.2). In particular, the intensity of the out of plane sodium mode at 277 cm$^{-1}$ in the Ni doped sample (Na(1)) reduces, whereas that of the 297 cm$^{-1}$ mode corresponding to Na(2) out of plane vibrations builds up (see inset a of Fig.2 and Fig.3a.). These observations clearly indicate that at low temperature Na(2) becomes the preferred site for occupation and this changeover in occupation occurs at 150 K which correlates with $T_{MIT}$ (Fig.3a.) In addition the constituent phonon modes exhibit significant changes in their line shapes associated with this transition (Fig.3). Whereas it is evident from inset (b) in Fig.2. that the corresponding sodium mode at 274 cm$^{-1}$ in the undoped sample follows a regular anharmonic temperature dependence. The evolution of the Na modes observed in $Na_{0.75}Co_{0.95}Ni_{0.05}O_2$ can be rationalized as discussed below. It has been conjectured [4,5] that in these systems charge ordering of the Co ions drives the system to the insulating state. While there is no direct evidence for the ordering of the $Co^{4+}$ ions, the induced Na ordering is believed to be a consequence of the former [4]. Hence in the present work we that the abrupt change in the intensity of the out of plane Na(1) and Na(2) modes along with the associated step like anomalies in both the line position and width provide definitive experimental evidence for the MIT being driven by $Co^{3+}/Co^{4+}$ charge ordering. It is also seen from Fig.2, that in addition to the changes in the sodium mode, the high frequency Co-O mode at ~570 cm$^{-1}$ exhibits a significant splitting on lowering the temperature for both $Na_{0.75}CoO_2$ and $Na_{0.75}Co_{0.95}Ni_{0.05}O_2$. Although such a splitting has earlier been attributed to the charge ordering in the $CoO_2$ planes [7], we are unable to rationalize this behavior on the same grounds due to the fact that a similar effect has also been observed for $Na_{0.75}CoO_2$- a system that remains metallic till the lowest of temperatures. Hence the origin of the splitting of the Co-O mode still remains unclear although one can always attribute such a splitting as arising due to a regular temperature effect that causes a better resolution of the phonons at low temperatures.

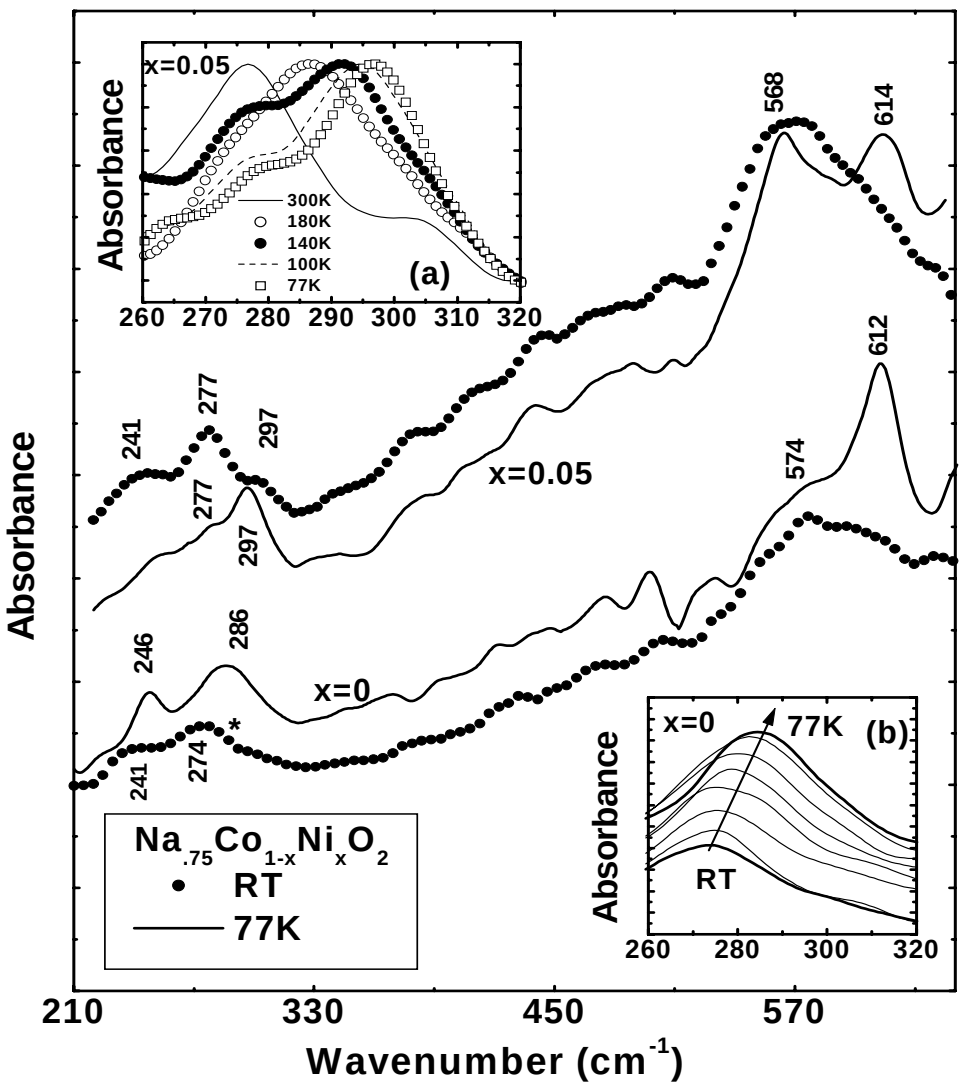

**Fig. 2** Room temperature and 77K FTIR absorption spectra of $Na_{.75}CoO_2$ and $Na_{0.75}Co_{0.95}Ni_{0.05}O_2$. Inset (a) shows the dramatic flipping of sodium mode at $T_{MIT}$ in the nickel doped sample while inset (b) shows a regular anharmonic hardening of the sodium mode due to temperature effects for the undoped sample.

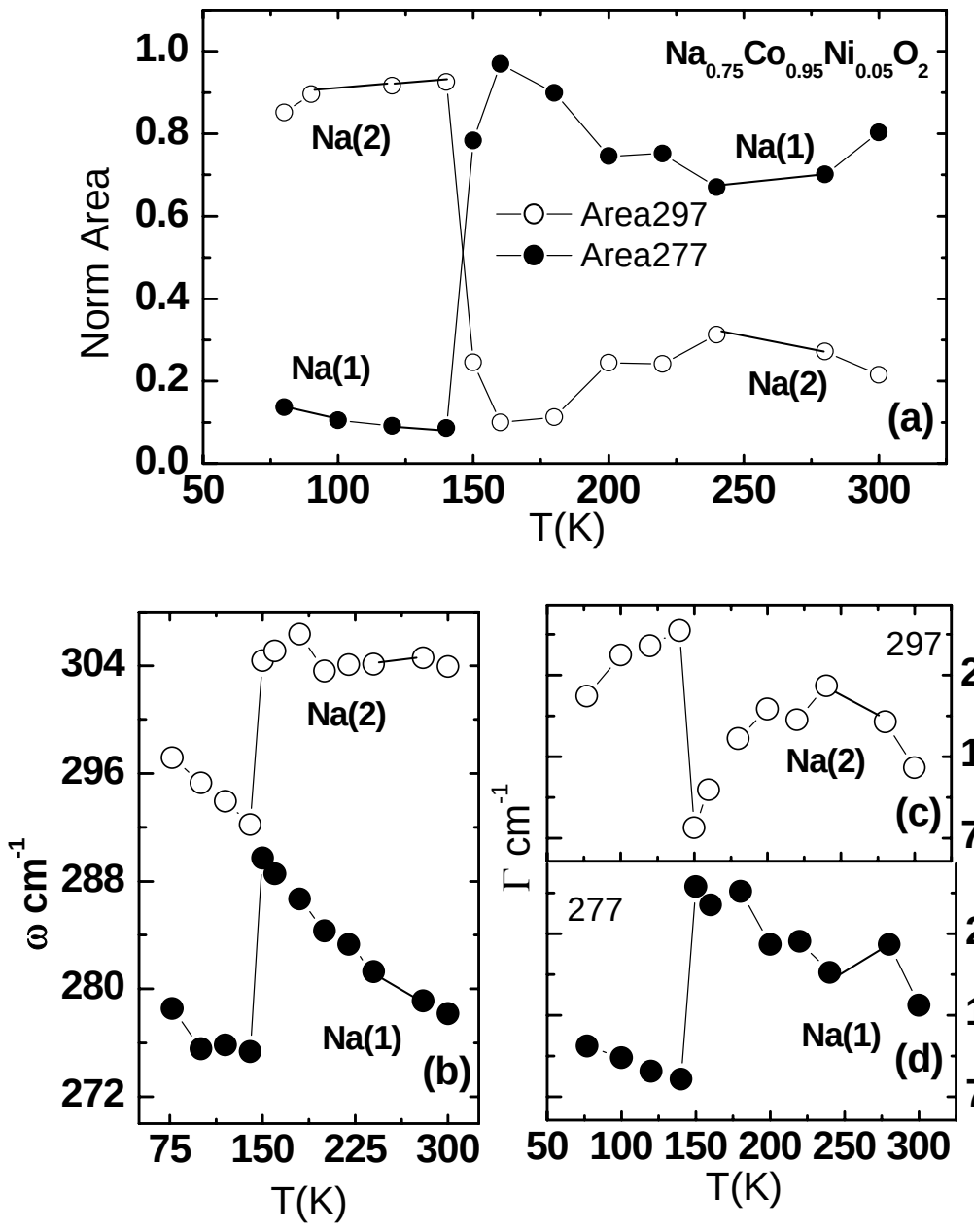

**Fig. 3**: (a) Dramatic flipping of the out of plane sodium mode intensities correlating with $T_{MIT}$. Sharp changes in line shape parameters: (b) phonon frequency corresponding to Na(1) and Na(2) sites, (c) & (d) phonon linewidths, that are seen to correlate with $T_{MIT}$ ~150K